# Treemble: A Graphical Tool to Generate Newick Strings from Phylogenetic Tree Images

John B. Allard[1,2]*, and Sudhir Kumar[1,2]

*Corresponding author

**Affiliations**

[1] Institute for Genomics and Evolutionary Medicine, Temple University, Philadelphia, PA 19122, USA

[2] Department of Biology, Temple University, Philadelphia, PA 19122, USA

**Abstract**

Phylogenetic trees are ubiquitous and central to biology, but most published trees are available only as visual diagrams and not in the machine-readable newick format. There are thus thousands of published trees in the scientific literature that are unavailable for follow-up analyses, comparisons, supertree construction, etc. Experts can easily read such diagrams, but the manual construction of a newick string is prohibitively laborious. Previous attempts to semi-automate the reading of tree images relied on image processing techniques. These quickly encounter difficulties with typical published tree diagrams that contain various graphical elements that overlap the branches, such as error bars on internal nodes. Here we introduce Treemble, a user-friendly desktop application for generating newick strings from tree images. The user simply clicks to mark node locations, and Treemble algorithmically assembles the tree from the node coordinates alone. Tip nodes can be automatically detected and marked. Treemble also facilitates the automatic reading of tip name labels and can handle both rectangular and circular trees. Treemble is a native desktop application for both MacOS and Windows, and is freely available and fully documented at treemble.org.

# 1 Introduction

Evolution is the single most important unifying principle in the life sciences, and as such understanding it is key to all biological research, either directly or indirectly (Dobzhansky, 1973). A crucial factor in understanding evolution is the ability to discern the ancestry and relationships of species. Tree diagrams (phylograms or cladograms) are the most common way to present inferred evolutionary relationships, whether derived from traditional morphological character-based analyses, population genetics, or molecular phylogenetics (Nei and Kumar, 2000). Vast numbers of such trees are published each year (the phrase "phylogenetic tree" produces over 2 million hits in Google Scholar). Phylogenetic tree figures typically communicate not only the inferred ancestral branching patterns (the tree topology), but also the branch lengths in units of the number of molecular substitutions per site, or may be time-calibrated and expressed in millions of years (Nei and Kumar, 2000).

These phylogenies are typically generated using data analysis software, which can export them in machine-readable newick format (Felsenstein, 1989) that encodes the topology and branch lengths. However, text files containing newick tree(s) are frequently missing from the supplementary information of publications that display phylogenetic tree diagrams. Therefore, we and other researchers wishing to use published phylogenies in downstream analyses need to translate graphical phylogeny displays into textual newick representations. However, generating a newick representation manually remains an arduous task, which requires hours for trees with even a small number of tips (<50) and days for bigger phylogenies.

For this reason, multiple software packages have been published to partially automate the process of acquiring a newick representation from a tree image (Hughes, 2011; Laubach and Haeseler, 2007; Laubach *et al.*, 2012). Both TreeSnatcher (Laubach and Haeseler, 2007; Laubach *et al.*, 2012) and TreeRipper (Hughes, 2011) detect tree branches in a figure by image processing techniques to identify the foreground in the form of contiguous dark pixels against a light background. This can be effective for optimally clean and annotation-free phylogenies, but the success rate for TreeRipper was low due to the complexity of many images, and this web application is no longer available at the published URL (Hughes, 2011). TreeSnatcher required the user to use image processing tools to condition the tree pixels to be detected, and in cases where extraneous intersecting lines, organismal silhouettes, and boxes are present, the user needed use tools to manually delineate the foreground to remove such distractions and make other manual modifications to aid in detection of nodes. In addition, entry of tip names required manually typing.

Here, we present a new software (Treemble) for building newick text strings from phylogenetic tree images semi-automatically and quickly. Treemble does not rely on image processing, so tree displays adorned with a multitude of annotations can be processed quickly. Treemble can handle images containing rectangular and circular phylogenies and measure branch lengths in the units of substitutions per site and time. In the following section, we describe Treemble's architecture, capabilities, and performance.

## 2 Results

Treemble provides a graphical interface for the user to click each node to mark it in the phylogeny display, which can be done in any order (Fig. 1a). We found this clicking procedure to take approximately one second per node, which means that nodes in a phylogeny with 50 tips can be marked within two minutes. This is much shorter than the time required to erase extraneous lines, boxes, and biological annotations that are often quite abundant in tree images. In addition, the tip nodes can be autodetected and marked in one step by drawing a rectangle encompassing them, which is particularly convenient for timetrees.

After the user finishes marking the nodes, Treemble automatically determines the connectivity of the nodes (see section 3 below) and displays overlays depicting the extracted tree (Fig. 1b). This display enables the user to verify that the phylogeny detected is correct. Treemble automatically highlights nodes that could not be fully connected, making it easy to find the site of any missing nodes. Of course, a user may choose to only mark-up a subset of nodes to produce a newick tree of a subset of taxa. The scale can also be calibrated based on a visual scale bar, and names of species can be imported from a text file which can be generated by the user's choice of optical character recognition software. We provide Tip Name Extractor GPT (https://chatgpt.com/g/g-rwiIPwboh-tip-name-extractor), which is a customized AI Chatbot designed to read species names from tree figures. It was created using OpenAI's GPT framework (OpenAI, 2023). A user simply drops their tree image into the AI and receives a link to download a text file which can be dragged and dropped onto the Treemble window to load tip names. Names are displayed next to each tip so spelling can be checked and an editor allows adjustments to be made within Treemble (Fig. 1b).

In addition to rectangular trees, Treemble has a specialized mode for extracting circular trees, including circular scale calibration and radial branch lengths, which no previous peer software provided (Fig. 1c). Polytomies and freeform connectivities can be specified by the user by manually connecting nodes to the correct parent with a simple two-click interaction. A cladogram mode allows any tree to be exported without branch lengths. Beyond simply capturing tree data, Treemble can also be used in a blank canvas mode with drawing tools, so users can quickly sketch a tree by hand and create a newick string according to their needs. An SVG image of any extracted tree can be exported directly. Full documentation with helpful visualizations is provided at Treemble.org, and a quickstart guide is available inside Treemble's interface.

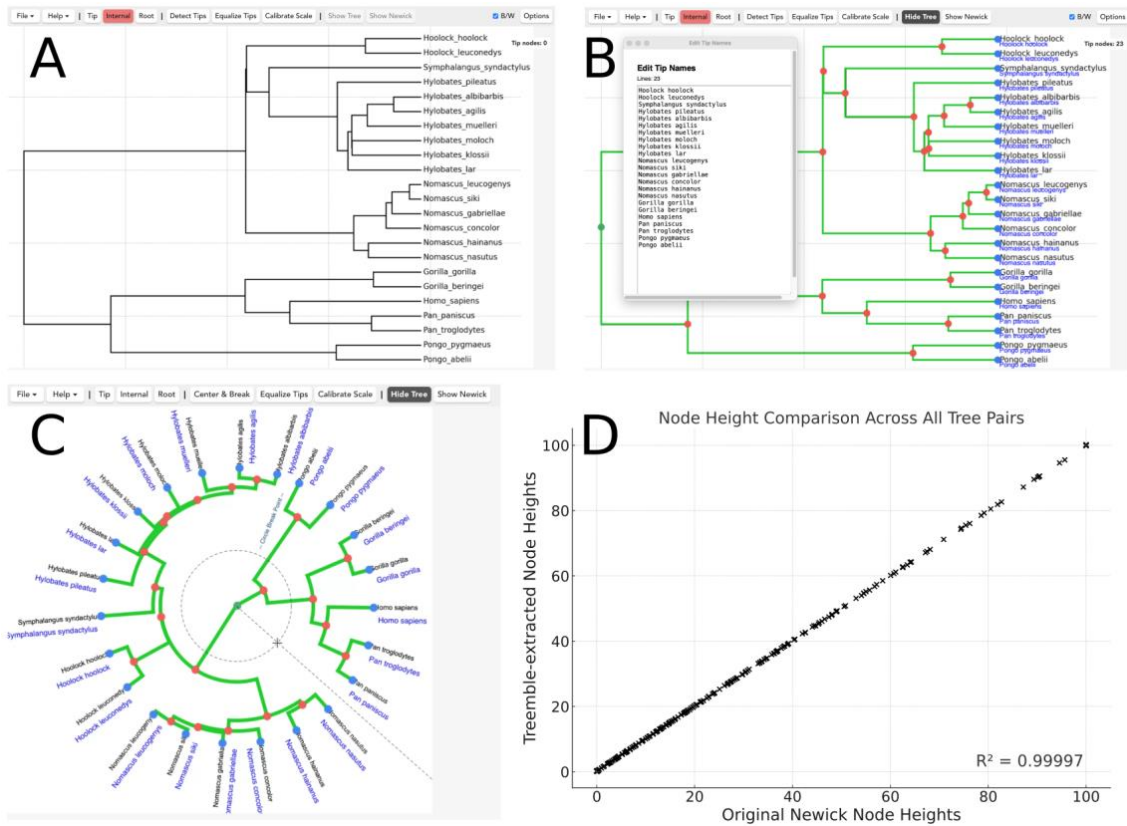

**Figure 1**. Treemble views and performance. **A)** The Treemble interface with a primate tree figure loaded. **B)** The treemble interface with tip and internal nodes marked and green branch overlays depicting the connectivity of the algorithmically reconstructed tree shown.  The tip name editor window is shown and name label overlays are shown in blue under each tip allowing the user to check the spelling of the labels. **C)** A circular tree is shown in the Treemble interface. The Show Tree option is On and green overlays overlap the tree image's branches.  **D)** A comparison of node heights from original simulated trees and trees extracted using Treemble from figures of the originals. The $R^2$ is 0.99997, indicating an extremely accurate recovery of branch lengths across these trials.

In order to test the accuracy of Treemble in recovering the branch lengths of trees from images, we generated ten simulated trees with 25 taxa each, using a lineage birth-death process as implemented in the dendropy package (Sukumaran and Holder, 2010). The resulting newick strings were rendered as tree images using biopython. Each image was loaded in Treemble which was used to export a Treemble-extracted newick string for the same tree. The resulting newick strings all perfectly matched the original topology. The node heights of each internal node were compared between the original and Treemble-extracted versions (Fig. 1d). The $R^2$ value was 0.99997 which indicates that the match is highly accurate. The total time taken to extract a newick string from each simulated tree image was 2 minutes on average. For a comparison of branch lengths, the $R^2$ value was 0.99994. The mean absolute error was 0.1118 with a mean branch length of 19.6067 (0.57%). This is very likely to be well within the phylogenetic reconstruction uncertainty. The trials were performed using relatively low-resolution images (width = 1,000 pixels), and high-resolution figures will lead to even better accuracy. All tree images, measurements, and code used to perform these simulations and trials are available in the supplementary dataset. In summary, Treemble is a highly accurate and reliable tool for recovering phylogenetic tree data from tree images.

**3 Implementation**

3.1 Software

Treemble is implemented as a native desktop application using the Tauri framework (v2.5.0). Tauri couples a small Rust-based backend with a system-native webview (WKWebView on macOS, WebView2 on Windows, and WebKitGTK on Linux), allowing the application logic and user interface to be written in TypeScript/React while retaining the size and performance characteristics of a compiled binary. The efficiency of the interface makes it possible to gather data from even very large trees, for instance, we completed a newick string from a phylogeny of 1,382 tips (Larridon *et al.*, 2021). Unlike Electron, where the runtime bundles an entire Chromium instance (resulting in binaries that are hundreds of megabytes), Tauri re-uses the host's browser engine and links statically to Rust libraries. As a result, Treemble installers for each system are smaller than 8 megabytes, so the software is fast to download and has a small footprint on a user's hard drive.

## 3.2 Tree assembly algorithm

Treemble uses a novel algorithm to assemble an adjacency graph based on $X, Y$ coordinates of nodes. The nodes are classified by the user as either tip or internal nodes. The algorithm iterates over the internal nodes from youngest to oldest (i.e., largest to smallest $X$ coordinate), connecting each one to the two younger parentless nodes that are nearest to it in the positive and negative $Y$ directions, respectively. A high level procedure follows:

1. Initialize the set of free nodes $F \leftarrow D$.
2. Collect the internal nodes
$$U = \{u \in D \mid t_u = \text{internal}\},$$
and sort them in strictly descending $x$.
3. For each internal node $u \in U$ in that order:
   1. Partition the current free set $F$ into
   $$F^+(u) = \{v \in F \mid x_v > x_u, y_v > y_u\},$$
   $$F^-(u) = \{v \in F \mid x_v > x_u, y_v < y_u\}.$$

   2. Choose
   $$v^+ = \arg\min_{v \in F^+(u)} (y_v - y_u)$$
   $$v^- = \arg\min_{v \in F^-(u)} (y_u - y_v)$$
   3. Attach edges $(u, v^+)$ and $(u, v^-)$ and remove $v^+, v^-$ from $F$.
4. Continue until $F = \emptyset$. A newick string can be obtained by recursion on the adjacency graph.
5. Branch lengths are obtained directly from the time axis:
$$L_{u \to v} = x_v - x_u.$$

## 3.3 Circular Trees

For circular trees, the X and Y dimensions are simply transformed into polar coordinates and the same algorithm applies, following the designation of a center point by the user.

Let the user–chosen center point be $(c_x, c_y)$, with a break angle $\theta_{\text{break}}$. Convert each screen coordinate $(x, y)$ to polar coordinates by

$$r = \sqrt{(x - c_x)^2 + (y - c_y)^2}, \qquad \theta = \text{atan2}(y - c_y, x - c_x) - \theta_{\text{break}}$$

Then the exact same algorithm from Section 3.2 applies, with

$$F^{\text{cw}}(u) = \{\, v \in F \mid r_v > r_u,\ 0 < (\theta_v - \theta_u) \bmod 2\pi < \pi \,\},$$
$$F^{\text{ccw}}(u) = \{\, v \in F \mid r_v > r_u,\ 0 < (\theta_u - \theta_v) \bmod 2\pi < \pi \,\}.$$

and choosing the nearest clockwise and counter-clockwise descendants by minimizing the angular difference. $\theta_{\text{break}}$ is used as the starting point for adding tip names labels.

## 4 Discussion

Treemble provides a solution to the problem of recovering machine-readable representations of phylogenetic trees from images, and it may also be useful for other fields where data are represented in hierarchical trees. A typical user can generate a newick string with Treemble in about 1 minute for every 10 taxa in a standard tree. Treemble has many quality-of-life features for users, including keyboard shortcuts, autosave and session recovery, tools for performing a diff analysis to compare two sets of tip names, automatic equalization of tip positions for ultrametric trees, grayscale mode to make seeing nodes easier, a range of visual options, and system dark mode support. The ability to save a csv file of node locations which can include tip names allows a Treemble session to be reopened at a later date and modified. In the future, the node location data paired with tree images may allow the training of a computer vision model to recognize internal nodes automatically, and Treemble makes gathering this data feasible. Treemble's simplicity and versatility will enable analyses of published phylogenies that were previously prohibitively difficult, paving the way for new advances in phylogenetics.

**Acknowledgements**

We thank Dr. Jack M. Craig, Kelly Abramowitz, Allen S. Thomas, Brandon K. Son, Ava Beasley, and Whitney L. Fisher for user feedback and suggestions. This work was supported by a fellowship to J.A. from Temple University and a grant from the National Science Foundation to S.K.

## Data and Code Availability

Installers for Treemble for MacOS and Windows are freely available for download at treemble.org, where full illustrated documentation can also be found.